\documentstyle[amssymb,aps]{revtex}

\input epsf

\begin{document}
\title{Resistive state of superconducting structures with fractal clusters of a
normal phase}
\author{Yu. I. Kuzmin}
\address{Ioffe Physical Technical Institute of the Russian Academy of Sciences,\\
Polytechnicheskaya 26 Street, Saint Petersburg 194021 Russia,\\
and Department of Physics, State Electrotechnical University,\\
Professor Popov 5 Street, Saint Petersburg 197376 Russia\\
e-mail: yurk@shuv.ioffe.rssi.ru; iourk@yandex.ru\\
tel.: +7 812 2479902; fax: +7 812 2471017}
\date{\today}
\maketitle

\begin{abstract}
The effect of morphologic factors on magnetic flux dynamics and critical
currents in percolative superconducting structures is considered. The
superconductor contains the fractal clusters of a normal phase, which act as
pinning centers. The properties of these clusters are analyzed in the
general case of gamma-distribution of their areas. The statistical
characteristics of the normal phase clusters are studied, the critical
current distribution is derived, and the dependencies of the main
statistical parameters on the fractal dimension are found. The effect of
fractal clusters of a normal phase on the electric field induced by the
motion of the magnetic flux after the vortices have been broken away from
pinning centers is considered. The voltage-current characteristics of
fractal superconducting structures in a resistive state for an
arbitrary fractal dimension are obtained. It is found that the fractality of
the boundaries of normal phase clusters intensifies magnetic flux trapping
and thereby increases the current-carrying capability of the superconductor.
\end{abstract}

\section{INTRODUCTION}

A significant property of clusters of a normal phase embedded in a
superconducting medium consists in their capability to trap a magnetic flux
and hold the vortices from moving under the action of the Lorentz force. As
a result, these clusters can act as effective pinning centers \cite{h1}- 
\cite{s4}. This property is widely used in creating composite
superconducting materials with high critical current values \cite{m5}, \cite
{b6}. The morphologic characteristics of the normal phase clusters 
essentially affects the dynamics of trapped magnetic flux, especially if the
clusters have fractal boundaries \cite{o7}-\cite{k9}. In the present work we
consider in detail the geometric probability properties of these fractal
clusters and their influence on the critical current and dynamics of trapped
magnetic flux near the transition of the superconductor into a resistive
state.

\section{FRACTAL GEOMETRY OF NORMAL PHASE CLUSTERS AND DISTRIBUTION OF
CRITICAL CURRENTS}

Let us consider a superconductor containing fragments of a normal phase.
Suppose that dimension of these fragments along one direction significantly
exceed the other sizes. Similar columnar defects are of great interest in
creating artificial pinning centers \cite{m5}, \cite{k10}-\cite{k12}. If
such a superconducting structure is cooled in a magnetic field directed
along the axis of alignment of these defects below the critical temperature,
then the distribution of magnetic flux trapped in the clusters of a normal
phase will be two-dimensional. This can be done especially easily with a
superconducting film where such clusters are formed near defects at the
boundary with the substrate during the growth process and are directed
transversely to the film plane \cite{m5}, \cite{k12}. Let us suppose that
the film surface fraction covered by the normal phase is below the
percolation threshold for the transfer of magnetic flux (i.e., less than
50\% for 2D-percolation \cite{s13}). In this case the fraction of the
superconducting phase exceeds the percolation threshold, so there is a
superconducting percolation cluster that can carry a transport current in
the film plane. This kind of structure provides for the effective pinning
and thereby raises the critical current, for the magnetic flux is trapped in
isolated clusters of the normal phase and the vortices cannot leave them
without crossing the surrounding superconducting space. As the current
increases, the trapped flux will remain unchanged until the vortices begin
to break away from those clusters for which the pinning force is less than
the Lorentz force caused by the transport current. When the magnetic flux is
breaking away from the pinning centers, each vortex must cross the infinite
superconducting cluster. In this case the vortices will move primarily along
the weak links that connect the clusters of a normal phase to each other 
\cite{d3}, \cite{b14}-\cite{h17}. These weak links form readily
in high-temperature superconductors (HTS) characterized by a small
coherence length \cite{m5}, \cite{s16}. Structural defects, which could act
simply as scattering centers at a large coherence length, create weak links
in HTS. There is a large variety of weak links in a wide range of spatial
scales in HTS \cite{m5}, \cite{b14}-\cite{k18}. At an atomic level weak
links are created by structural point defects, primarily by oxygen vacancies 
\cite{s16}, \cite{r19}. On a mesoscopic scale weak links are efficiently
formed on twin boundaries \cite{n15}, \cite{s16}, \cite{k18}, \cite{m20}.
Twins can be located at a distance of several tens of nanometers from each
other, therefore even a single crystal can have a fine substructure created
by twins. Finally, on a macroscopic scale weak links are created by various
structural defects, such as boundaries of grains and crystallites or
barriers arising from the secondary degrading of a non-stoichiometric crystal
into the domains with a high and low oxygen content \cite{h17}-\cite{r19}.
Moreover, a magnetic field favors the coherence length to further decrease 
\cite{s21}, so weak link formation is facilitated still more. In
conventional low-temperature superconductors, which are of large coherence
length, weak links are formed due to the proximity effect at sites where the
distance between neighboring clusters of a normal phase is minimum.

Thus, regardless of their origin, weak links form channels for vortex
transport where the pinning force is less than the Lorentz force caused by
the transport current. According to their configuration each cluster of the
normal phase has its own value of the depinning current that makes the
contribution to the overall statistical distribution of the critical
currents. When a transport current is passed through the sample, vortices
will be broken away first from clusters of a smaller pinning force and,
accordingly, a smaller critical current of depinning. So the change in the
trapped magnetic flux will be proportional to the number of the normal phase
clusters which have the critical current less than the preset value $I$.
Therefore, the relative decrease in the flux can be expressed through the
function of cumulative probability $F=F\left( I\right) $ for the
distribution of critical currents of the clusters: 
\begin{equation}
\frac{\Delta \Phi }{\Phi }=F\left( I\right)   \label{eq1}
\end{equation}
where 
\[
F\left( I\right) =\Pr \left\{ \forall I_{j}<I\right\} 
\]

The right-hand side of Eq.~(\ref{eq1}) is the probability that any $j$-th
cluster has a the critical current $I_{j}$ less than the given upper bound $I
$. On the other hand, the magnetic flux trapped in a single cluster is
proportional to its area $A$. Therefore, the decrease in the trapped flux
can be expressed through the function of cumulative probability $W=W\left(
A\right) $ for the distribution of the areas of normal phase clusters, which
is proportional to the number of clusters of area smaller than a given value
of $A$%
\begin{equation}
\frac{\Delta \Phi }{\Phi }=1-W\left( A\right)   \label{eq2}
\end{equation}
where 
\[
W\left( A\right) =\Pr \left\{ \forall A_{j}<A\right\} 
\]

Generally, the distribution of cluster areas can be described by the
gamma-distribution, for which the function of cumulative probability has the
form 
\begin{equation}
W(A)=%
{\displaystyle{\gamma (g+1,A/A_{0}) \over \Gamma (g+1)}}%
\label{eq3}
\end{equation}
where $\gamma \left( \nu ,z\right) $ is the incomplete gamma function, $%
\Gamma \left( \nu \right) $ is Euler gamma function, and $A_{0}$ and $g$ are
the parameters of the gamma-distribution that determine the mean area of
the normal phase cluster $\overline{A}=(g+1)A_{0}$ and its standard
deviation $\sigma _{A}=A_{0}\sqrt{g+1}$. The distribution function of
cluster areas can be found from the geometric probability analysis of
electron photomicrographs of superconducting films \cite{k8}, \cite{k11}, 
\cite{k12}. For example, in the practically important case of YBCO films
with columnar defects \cite{k12} the exponential distribution is realized,
which is a special case of the gamma-distribution of Eq.~(\ref{eq3}) with $%
g=0$.

To clarify how the transport current affects the trapped magnetic flux, it
is necessary to find the relationship between the distribution of critical
currents of clusters of Eq.~(\ref{eq1}) and the distribution of their areas
of Eq.~(\ref{eq2}). The larger the normal phase cluster size, the more weak
links rise from its perimeter bordering with the surrounding superconducting
space, and therefore, the smaller is the critical current at which the
magnetic flux breaks away from this cluster. Let us suppose that the weak
link concentration per unit length of the perimeter is the same for all
clusters and all the clusters of equal perimeters have the same pinning
force. Then the critical current $I$ is inversely proportional to the
perimeter $P$ of the normal phase cluster: $I\propto 1/P$, because the
probability to find a weak link over the larger perimeter is higher. It
is suggested here that the vortex, driven by the Lorentz force, passes
through the weak link from one normal phase cluster to another with the
probability of 100\%. In this case magnetic flux is transferred through the
superconducting medium by Josephson vortices. The probability to trap the
vortex when it moves through the weak link under the action of the Lorentz
force is negligible, inasmuch as the Josephson penetration depth in the
considered materials significantly exceeds the sizes of possible
irregularities along the weak link. In accordance with these assumptions, in
order to find the distribution function of Eq.~(\ref{eq1}) we have to know
the relationship between the perimeter and the area of a cluster. As was
first established in Ref.~\cite{k8}, fractal properties of the boundaries of
normal phase clusters essentially affect the dynamics of magnetic flux in
superconductors. For fractal clusters the perimeter-area relation has the
form \cite{m22} 
\begin{equation}
P\propto A^{D/2}  \label{eq4}
\end{equation}
where $D$ is the fractal dimension of the cluster boundary.

Relation of Eq.~(\ref{eq4}) agrees with the generalized Euclid theorem \cite
{m23}, according to that the ratios of the corresponding measures are equal
when they are reduced to the same dimension. So $P^{1/D}\propto A^{1/2}$,
and this relation holds true both for Euclidean clusters (for which the
Hausdorff--Bezikovich dimension of the perimeter is equal to the topological
dimension of a line, $D=1$), and for fractal clusters (for which the
Hausdorff--Bezikovich dimension of the boundary strictly exceeds the
topological one, $D>1$).

Note that it is just the statistical distribution of cluster areas, rather
than their perimeters, is fundamental for deriving the critical current
distribution. Inasmuch as the Hausdorff--Bezikovich dimension of a fractal
line exceeds unity, the perimeter of a fractal cluster is not well defined:
it diverges when the precision of its measurement increases indefinitely 
\cite{m22}. At the same time, the topological dimension of the cluster area
coincides with the Hausdorff--Bezikovich dimension (both are equal to 2).
Therefore, the area of a surface confined by the fractal curve is a finite,
well-defined quantity.

Analyzing the geometric characteristics of normal phase clusters, we are
considering the cross-section of extended columnar defect by the plane the
transport current flows through. Therefore, though a normal phase cluster is
a self-affine fractal \cite{m24}, we can consider its geometric probability
properties in the planar section only, where the boundary of cluster is
statistically self-similar.

Now, using relation of Eq.~(\ref{eq4}) between the fractal perimeter and
area, together with the assumption of inverse proportionality of the
critical current of a cluster to its perimeter, we arrive at the following
expression for critical current of the cluster: $I=\alpha A^{-D/2}$%
, where $\alpha $ is a form factor. Taking into account our initial
relations of Eqs.~(\ref{eq1}) and (\ref{eq2}), we can find the distribution
of critical currents in the general case of the gamma-distributed cluster
areas: 
\begin{equation}
F(i)=\frac{\Gamma (g+1,Gi^{-2/D})}{\Gamma (g+1)}  \label{eq5}
\end{equation}
where 
\[
G\equiv \left( \frac{\theta ^{\theta }}{\theta ^{g+1}-\left( D/2\right) \exp
\left( \theta \right) \Gamma (g+1,\theta )}\right) ^{\frac{2}{D}}\text{\ \ \
\ , \ \ \ \ \ \ \ \ \ \ \ \ \ \ \ \ \ \ \ \ \ \ \ \ \ }\theta \equiv g+1+%
\frac{D}{2} 
\]
$\Gamma \left( \nu ,z\right) $ is the complementary incomplete gamma
function, $i\equiv I/I_{c}$ is the dimensionless electric current, and $%
I_{c}=\alpha \left( A_{0}G\right) ^{-D/2}$ is the critical current of the
transition into a resistive state. The found function of cumulative
probability of Eq.~(\ref{eq5}) provides a complete description of the effect
of the transport current on the trapped magnetic flux. Using this function,
we can easily derive the probability density $f\left( i\right) =dF/di$ for
the distribution of depinning currents, which has the form 
\begin{equation}
f(i)=\frac{2G^{g+1}}{D\Gamma (g+1)}i^{-\frac{2}{D}(g+1)-1}\exp \left( -Gi^{-%
\frac{2}{D}}\right)  \label{eq6}
\end{equation}

The probability density is normalized to unity over all possible positive
values of the critical currents. The relative change in the trapped flux $%
\Delta \Phi /\Phi $, which can be directly found from Equation (\ref{eq5}),
also determines the density of vortices $n$ broken away from pinning centers
by the current of a given magnitude $i$%
\begin{equation}
n\left( i\right) =\frac{B}{\Phi _{0}}\int_{0}^{i}f\left( i^{\prime }\right)
di^{\prime }=\frac{B}{\Phi _{0}}\frac{\Delta \Phi }{\Phi }  \label{eq7}
\end{equation}
where $B$ is the magnetic field, and $\Phi _{0}\equiv hc/\left( 2e\right) $
is the magnetic flux quantum ($h$ is Planck constant, $c$ is the velocity of
light, and $e$ is the electron charge).

Figure \ref{figure1} shows how $g$-parameter of the gamma-distribution
affects the distribution of the critical currents of the clusters. As an
example, we have taken a value of the fractal dimension $D=1.5$, which is
near to the value $D=1.44\pm 0.02$ found earlier \cite{k8} for the normal
phase clusters in YBCO film structures. On the other hand, the value $D=1.5$
is intermediate between two limiting cases: $D=1$ for the Euclidean clusters
and $D=2$ for the clusters of the most fractality. When the fractal
dimension differs from unity so much, the fractal properties of the cluster
structure of a superconductor are of great importance. In Fig.~\ref{figure2}
the cluster area distribution is presented. The corresponding probability
density has the following form 
\begin{equation}
w\left( a\right) =\frac{\left( g+1\right) ^{g+1}}{\Gamma \left( g+1\right) }%
a^{g}e^{-\left( g+1\right) a}  \label{eq8}
\end{equation}
where $a\equiv A/\overline{A}$ is the dimensionless cluster area, with $%
\overline{a}=1$ and $\sigma _{a}=1/\sqrt{g+1}$. The function of cumulative
probability for the dimensionless area is related with function of Eq.~(\ref
{eq8}) by the formula $W\left( a\right) =\int\nolimits_{0}^{a}w\left(
a^{\prime }\right) da^{\prime }$ and can be written as 
\[
W\left( a\right) =\frac{\gamma \left( \left( g+1,(g+1\right) a\right) }{%
\Gamma \left( g+1\right) } 
\]

As is seen from Fig.$~$\ref{figure1}, when $g$-parameter decreases, the
distribution $f=f\left( i\right) $ spreads to the right, spanning higher and
higher range of critical currents. It might be expected that the highest
current-carrying capability of the superconductor can be achieved in the
limiting case of $g=0$, when the gamma-distribution of Eq.~(\ref{eq8}) is
reduced to the exponential one $w\left( a\right) =\exp \left( -a\right) $.
In this case the small clusters give the most contribution into the overall
distribution (curve 1 in Fig.~\ref{figure2}). For $g=0$ expressions of Eqs.~(%
\ref{eq5}) and (\ref{eq6}) can be simplified 
\begin{equation}
F\left( i\right) =\exp \left( -\left( \frac{2+D}{2}\right) ^{\frac{2}{D}%
+1}i^{-\frac{2}{D}}\right)  \label{eq9}
\end{equation}

\begin{equation}
f\left( i\right) =\frac{2}{D}\left( \frac{2+D}{2}\right) ^{\frac{2}{D}+1}i^{-%
\frac{2}{D}-1}\exp \left( -\left( \frac{2+D}{2}\right) ^{\frac{2}{D}+1}i^{-%
\frac{2}{D}}\right)  \label{eq10}
\end{equation}
where, as above, $i\equiv I/I_{c}$, and the critical current of the
resistive transition can be calculated using a simpler formula: $%
I_{c}=\left( 2/\left( 2+D\right) \right) ^{\left( 2+D\right) /2}\alpha
A_{0}^{-D/2}$.

Figure \ref{figure3} shows how the fractal dimension of the cluster
boundaries affects the distribution of the critical currents of
Eq.~(\ref{eq10}). As is clearly seen from the figure, with increasing
fractal dimension the critical current distribution spreads and shifts to
the higher values of current. This shift can be characterized by
the dependence of the mean critical current on the fractal dimension, as
is shown in Fig.~\ref{figure4}. Although the mode of the distribution
{\it mode}$f\left( i\right) =\left( 2+D\right) /2$\ 
is linear with respect to the fractal
dimension, the mathematical expectation obeys a much stronger superlinear
law controlled by Euler gamma function 
\begin{equation}
\overline{i}=\left( \frac{2+D}{2}\right) ^{\frac{2+D}{2}}\Gamma \left( 1-%
\frac{D}{2}\right)  \label{eq11}
\end{equation}

Thus, the geometric probability properties of the normal phase clusters
determine the statistical distribution of depinning currents. Using this
distribution along with Equation (\ref{eq1}), the relative change in the
trapped flux under the action of the transport current can be found.

\section{DYNAMICS OF MAGNETIC FLUX TRAPPED IN NORMAL PHASE CLUSTERS WITH
FRACTAL BOUNDARIES}

The effect of the transport current on the trapped magnetic flux is
demonstrated in Fig.~\ref{figure5}. The change in the trapped flux was
calculated using formula of Eq.~(\ref{eq1}) for the exponential-hyperbolic
critical current distribution of Eq.~(\ref{eq9}). Curves 1 and 5 in Fig.~\ref
{figure5} correspond to the limiting cases of the Euclidean clusters ($D=1$)
and the clusters of maximum fractality ($D=2$), respectively; they confine
the region of changes in the trapped flux for all possible values of the
fractal dimension. The function $F=F\left( i\right) $ reveals one important
property; namely, it is very flat near the coordinate origin. It can readily
be shown that all its derivatives become zero at the origin point: $%
d^{k}F\left( 0\right) /di^{k}=0$ for any values of $k$. Therefore even the
Taylor series expansion of the function in the vicinity of the origin
converges to zero value but not to the quantity $F$ itself. This
mathematical feature has a clear physical sense: such a small transport
current does not affect the trapped magnetic flux, because there are no
pinning centers with such small critical currents in the total statistical
distribution. The decrease in the magnetic flux becomes noticeable only near
the point of a resistive transition ($i=1$). A use of the exponential-hyperbolic
distribution of critical currents of Eq. (9) excludes the uncertainty caused
by truncation of non-physical negative values of the depinning currents,
which may happen, e.g., in the case of the normal distribution \cite{k25}, 
\cite{b27}.

\begin{figure}[tbp]
\epsfbox{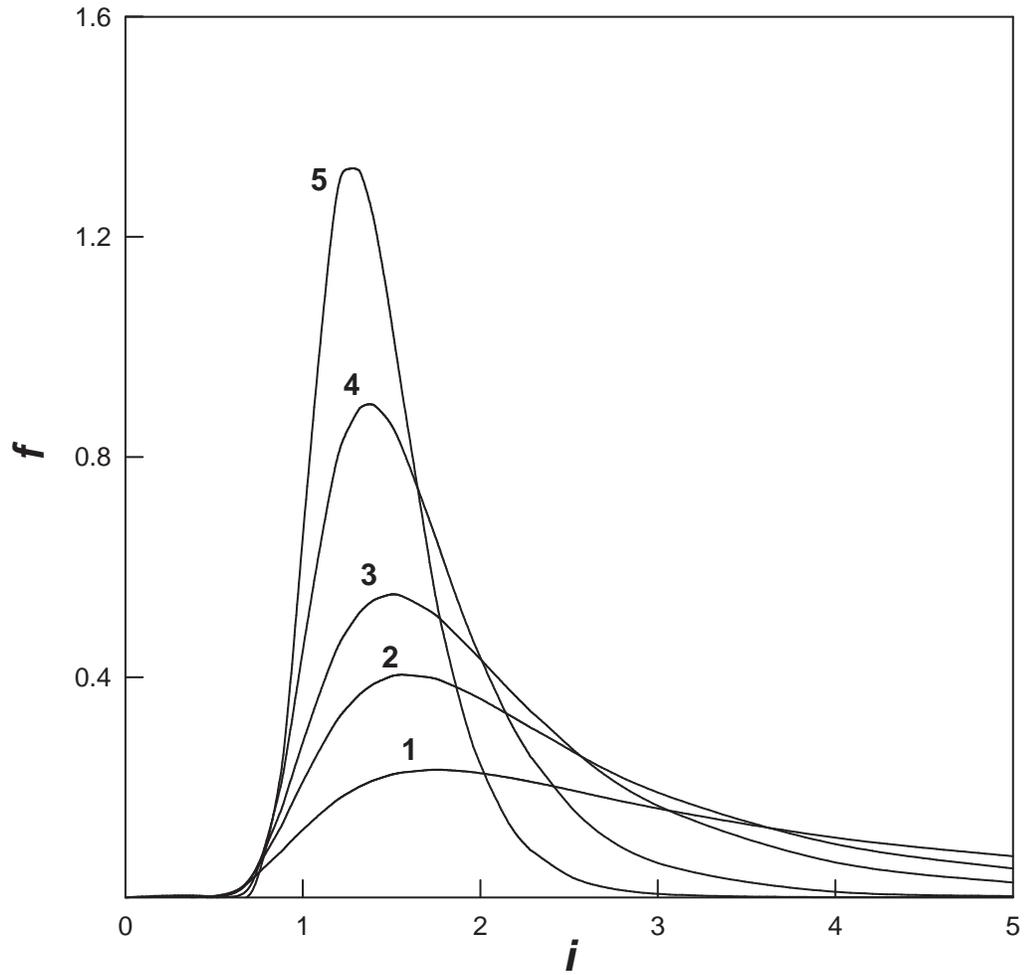}
\caption{Distribution of the critical currents for various values of $g$%
-parameter of the gamma-distribution at the fractal dimension of $D=1.5$.
Curve (1) corresponds to the case of $g=0$, (2) - $g=1$, (3) - $g=2$,
(4) - $g=5$, and (5) - $g=10$.}
\label{figure1}
\end{figure}

\begin{figure}[tbp]
\epsfbox{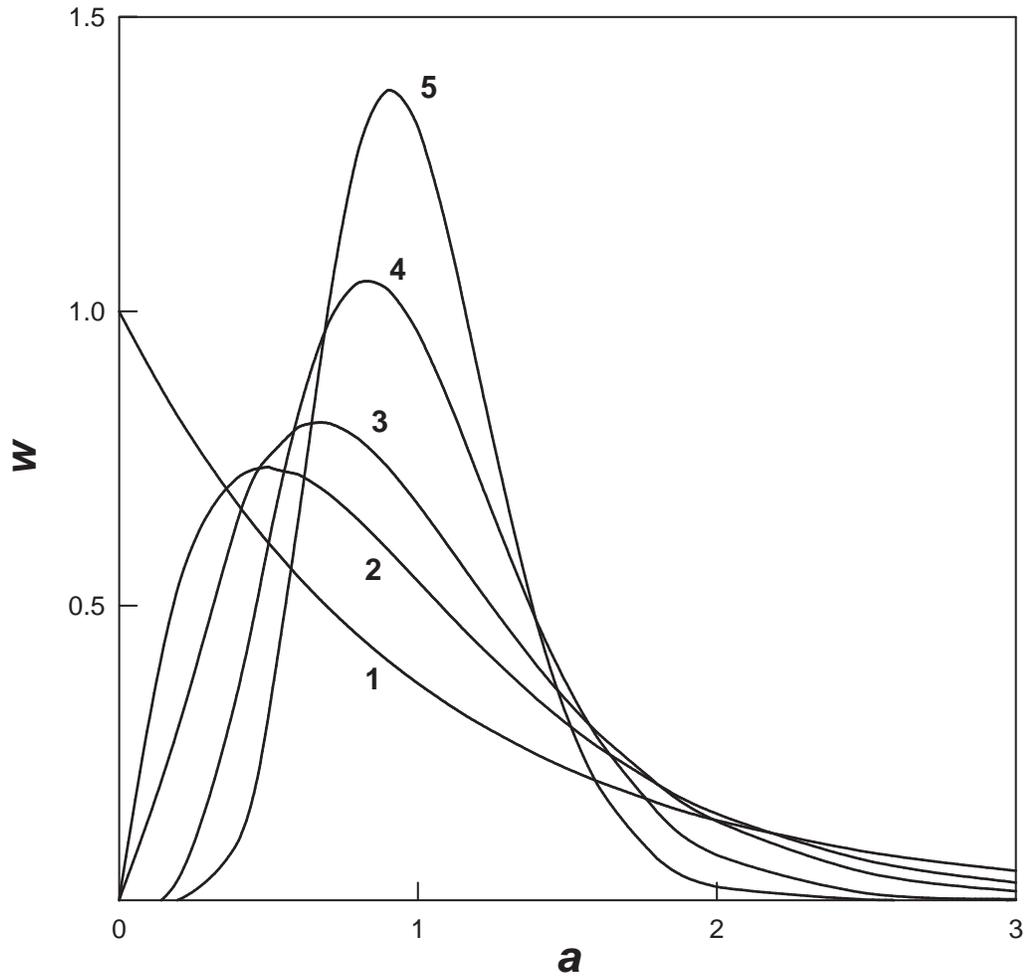}
\caption{Gamma-distribution of the areas of normal phase clusters. Curve (1)
corresponds to $g=0$, (2) - $g=1$, (3) - $g=2$, (4) - $g=5$, and (5) - $g=10$%
.}
\label{figure2}
\end{figure}

\begin{figure}[tbp]
\epsfbox{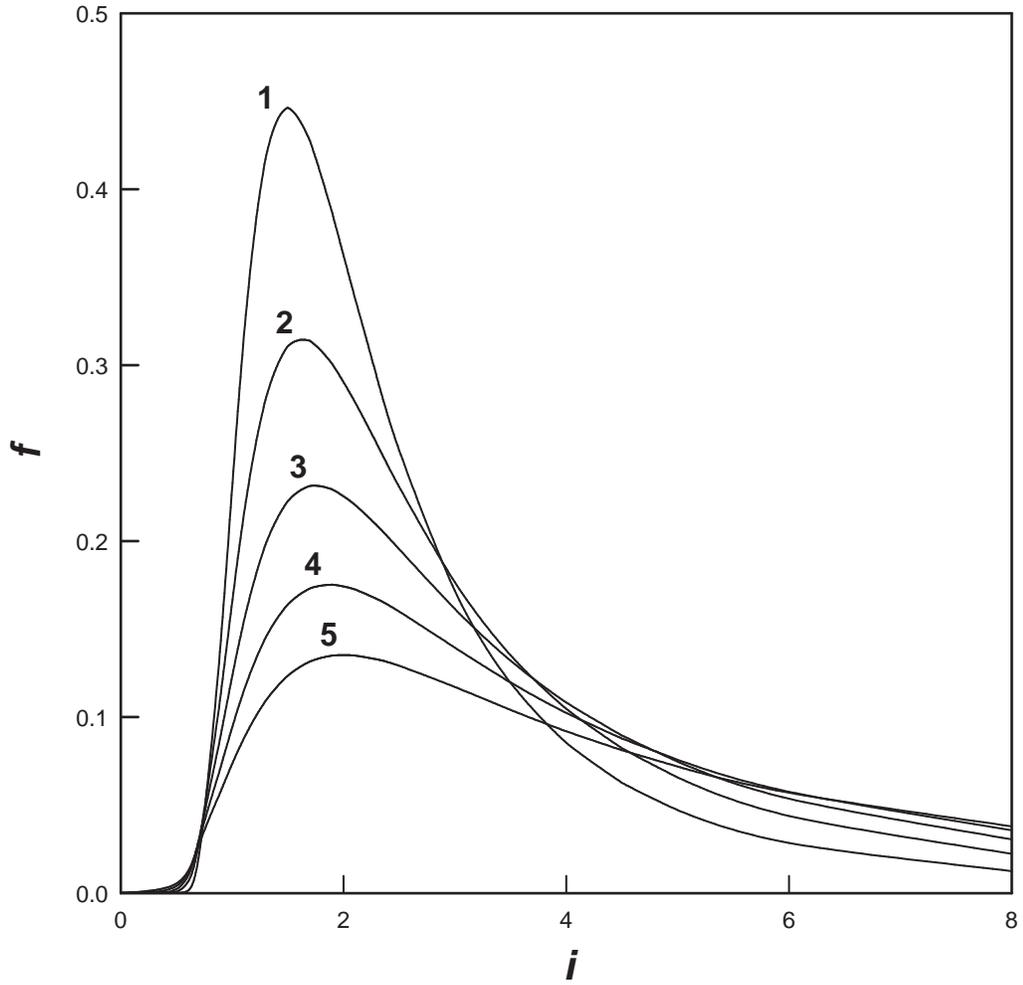}
\caption{Effect of the fractal dimension of the cluster boundary on the
critical current distribution at $g=0$. Curve (1) corresponds to the case of
Euclidean clusters ($D=1$), curve (2) - to the clusters of fractal dimension 
$D=1.25$, (3) - $D=1.5$, (4) - $D=1.75$, and (5) - to the clusters of
boundaries with maximum fractality ($D=2$).}
\label{figure3}
\end{figure}

As is seen from Fig.~\ref{figure5}, the breaking of the vortices away is
observed mainly at $i>1$, when the sample undergoes a transition into a
resistive state. Figure \ref{figure5} demonstrates the practically important
property of a superconducting structure with fractal normal phase clusters:
the fractality favors magnetic flux trapping, and thereby increases the
critical current magnitude below which the sample remains in a
superconducting state. Indeed, the transport current of the value $i=2$
causes the 43\% of the total trapped magnetic flux to break away from the
usual Euclidean clusters ($D=1$, curve 1), whereas this value equals only
to 13.5\% for the fractal normal phase clusters of the greatest possible
fractal dimension ($D=2$, curve 5). This is equivalent to the pinning
increase of 218\% in the latter case. The enhancement of pinning due to
fractality can be characterized by the pinning gain factor 
\[
k_{\Phi }\equiv 20\log \frac{\Delta \Phi \left( D=1\right) }{\Delta \Phi
\left( D\right) }\text{, \ \ dB} 
\]
which is equal to the relative decrease in the fraction of the magnetic flux
broken away from the fractal clusters of dimension $D$ in comparison with
the Euclidean ones ($D=1$). In Fig.$~$\ref{figure6} the dependencies of the
pinning gain factor on the transport current and on the fractal dimension
are presented. The maximum pinning gain is achieved when the boundaries of
clusters have the greatest possible fractal dimension ($D=2$). Note that the
pinning gain factor characterizes the properties of the superconductor in 
the range of transport currents corresponding to a resistive state ($i>1$).
With smaller currents, the trapped magnetic flux virtually does not change,
because there are no pinning centers with such small critical currents
(Figs.~\ref{figure1}, \ref{figure3}) and the breaking of the vortices away
has not started yet. In the presence of a finite resistance any current flow
is accompanied by energy dissipation. As for any hard superconductor
(type-II, with pinning centers) the energy dissipation in a resistive state
does not mean the phase coherence destruction yet. Some dissipation
accompanies any motion of the magnetic flux; this effect can be observed in
a hard superconductor even for small transport currents. Therefore, the 
critical current in such materials cannot be determined as the greatest
dissipationless current. The superconducting state collapses only when the
dissipation increases in an avalanche-like manner as a result of the
thermo-magnetic instability.

\begin{figure}[tbp]
\epsfbox{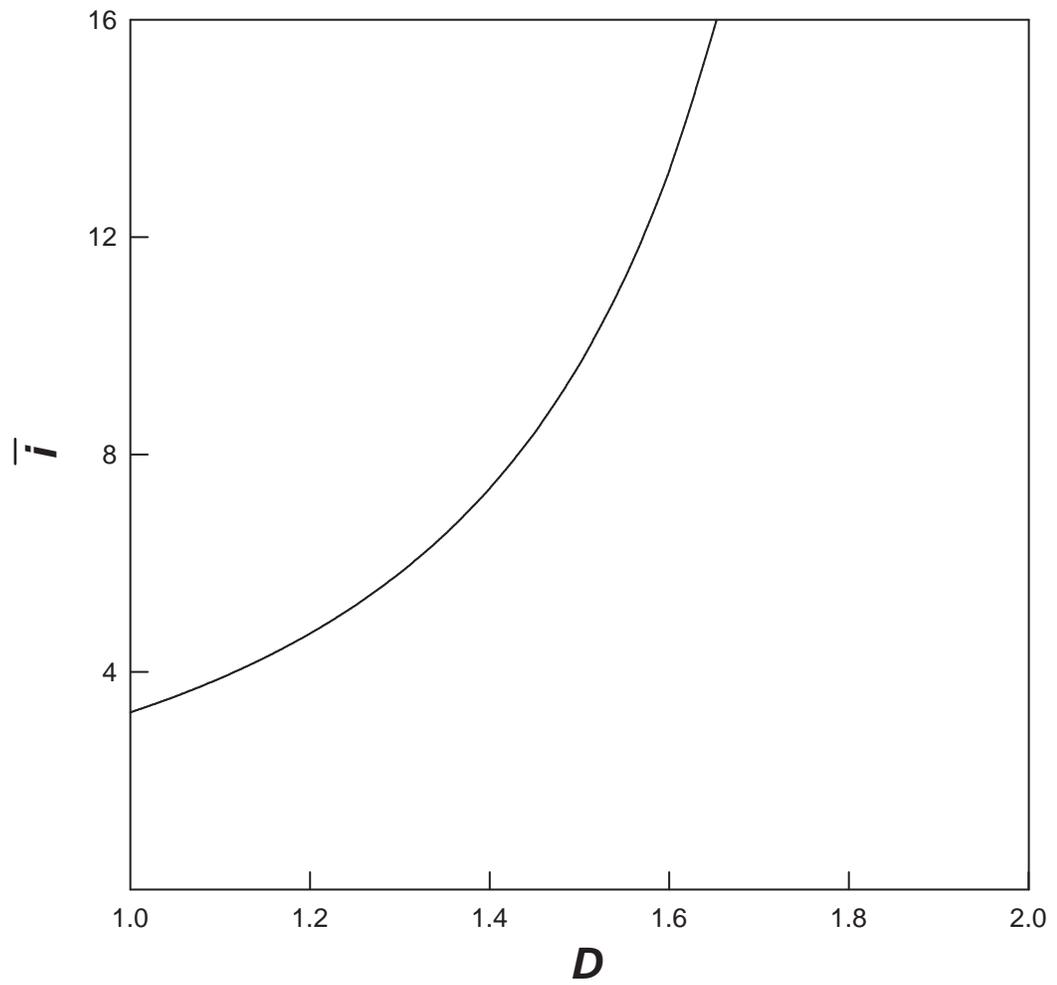}
\caption{Dependence of the mean critical current on the fractal dimension.}
\label{figure4}
\end{figure}

\begin{figure}[tbp]
\epsfbox{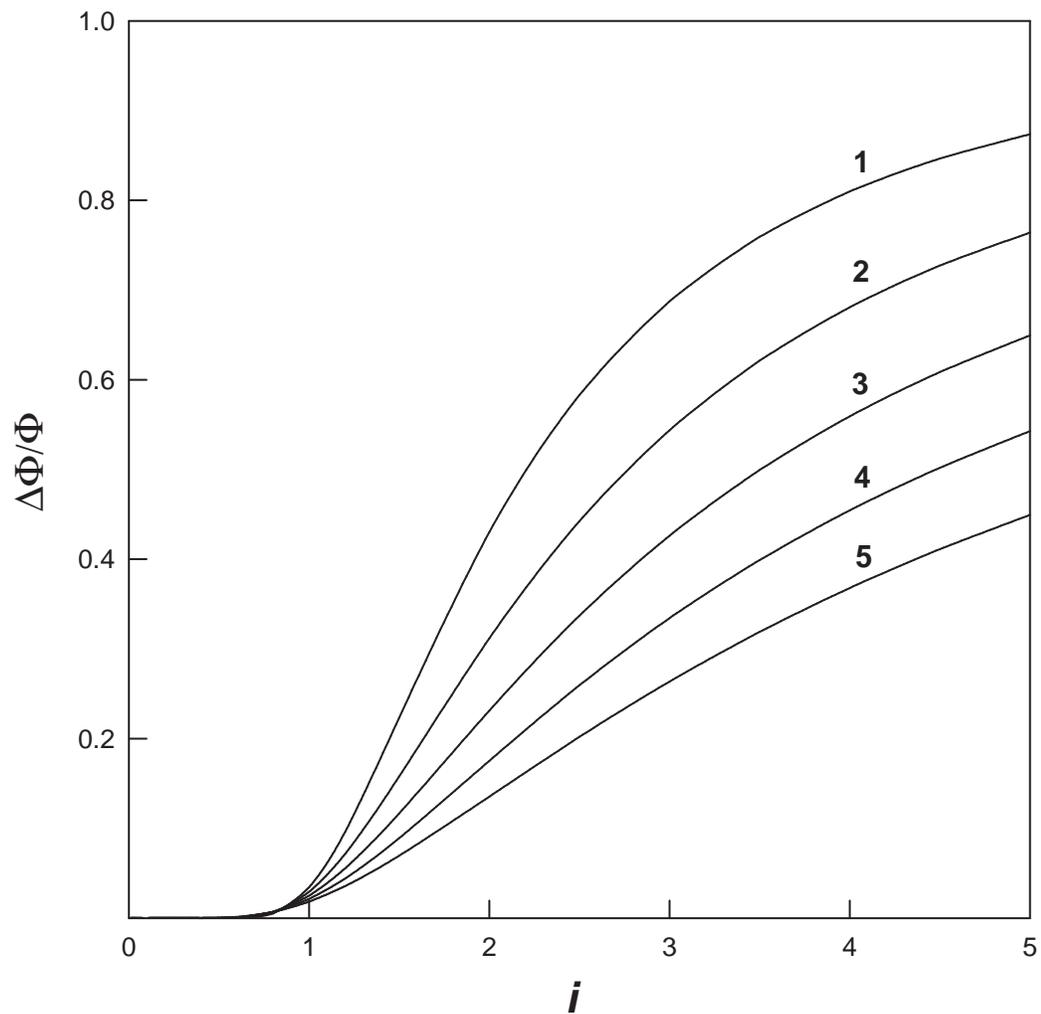}
\caption{Effect of a transport current on the magnetic flux trapped in
fractal clusters of a normal phase at $g=0$. Curve (1) corresponds to $D=1$,
(2) - $D=1.25$, (3) - $D=1.5$, (4) - $D=1.75$, (5) - $D=2$.}
\label{figure5}
\end{figure}

\begin{figure}[tbp]
\epsfbox{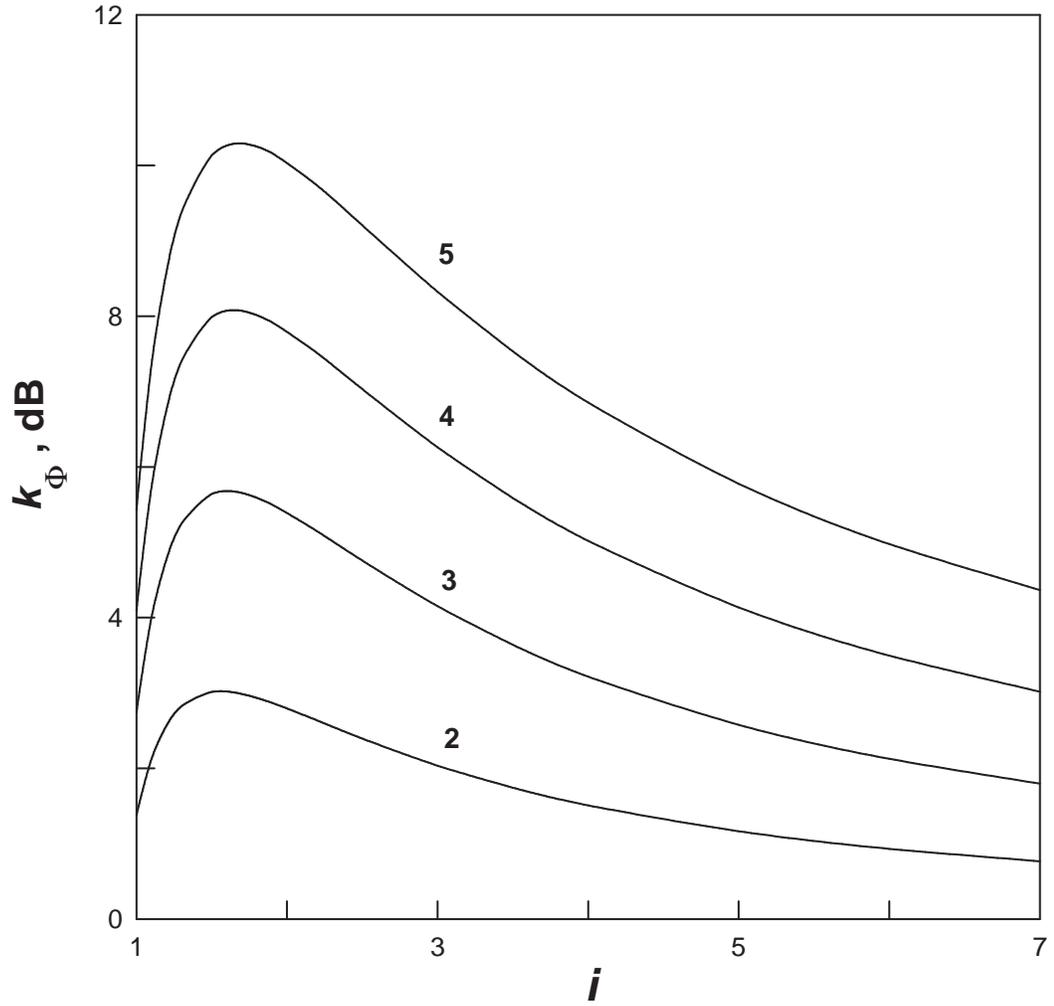}
\caption{The pinning gain factor at different values of the fractal
dimension. Curve (2) corresponds to $D=1.25$, (3)
- $D=1.5$, (4) - $D=1.75$, (5) - $D=2$.}
\label{figure6}
\end{figure}

The reason for the pinning gain caused by the fractality of the cluster
boundaries lies in the fundamental properties of the statistical
distribution of critical currents (see Fig.~\ref{figure4}). In the case of
Euclidean clusters the mean value of the critical current calculated in
accordance with Eq.~(\ref{eq11}) is equal to $\overline{i}\left( D=1\right)
=(3/2)^{3/2}\sqrt{\pi }=3.2562$, whereas for the clusters of the most
fractality ($D=2$) it is infinitely high. As is seen from Fig.~\ref{figure3}%
, when the fractal dimension increases, the contribution of the clusters
with high depinning currents to the overall statistical distribution
increases, too, which results in the enhancement of magnetic flux trapping.

In a resistive state a hard superconductor can be adequately described by
its voltage-current characteristics. Using the fractal distribution of the
critical currents of Eq.~(\ref{eq6}), we can find the electric field caused
by magnetic flux motion after the vortices have been broken away from
pinning centers. Inasmuch as each cluster of the normal phase contributes to
the overall distribution of the critical currents, the voltage across the
superconductor $V=V\left( i\right) $ is a response to the sum of effects
made by contribution from each cluster. This response can be expressed as a
convolution integral 
\begin{equation}
V=R_{f}\int\limits_{0}^{i}(i-i^{\prime })f(i^{\prime })di^{\prime }
\label{eq12}
\end{equation}
where $R_{f}$ is the flux flow resistance. This representation for the
voltage across the sample is often used to consider the pinning of bunches
of vortex lines in a superconductor \cite{w28}, as well as to analyze the
critical scaling of voltage-current characteristics \cite{b27}, i.e., in all
the cases when the distribution of depinning critical currents occurs. In
the following analysis we will focus on the results of the properties of the
exponential-hyperbolic distribution in Eq.~(\ref{eq10}); we will not
consider questions associated with the possible dependence of the flux flow
resistance $R_{f}$ on the transport current.

In the simplest case, when all pinning centers have the same critical
current $i_{c}$, all vortices are released simultaneously at $i=i_{c}$, and
their density, in accordance with Eq.~(\ref{eq7}), has the form 
\[
n=\frac{B}{\Phi _{0}}\int_{0}^{i}\delta \left( i^{\prime }-i_{c}\right)
di^{\prime }=\frac{B}{\Phi _{0}}h\left( i-i_{c}\right) 
\]
where $\delta \left( i\right) $ is the Dirac delta function, and $h\left(
i\right) \equiv \left\{ 
\begin{array}{cc}
1 & \text{\ \ \ for \ \ }i\geqslant 0 \\ 
0 & \text{ \ \ for \ \ }i<0
\end{array}
\right. $\ is the Heaviside function. The trapped flux would change at once
by 100\% in this case: $\Delta \Phi /\Phi =h\left( i-i_{c}\right) $.

\begin{figure}[tbp]
\epsfbox{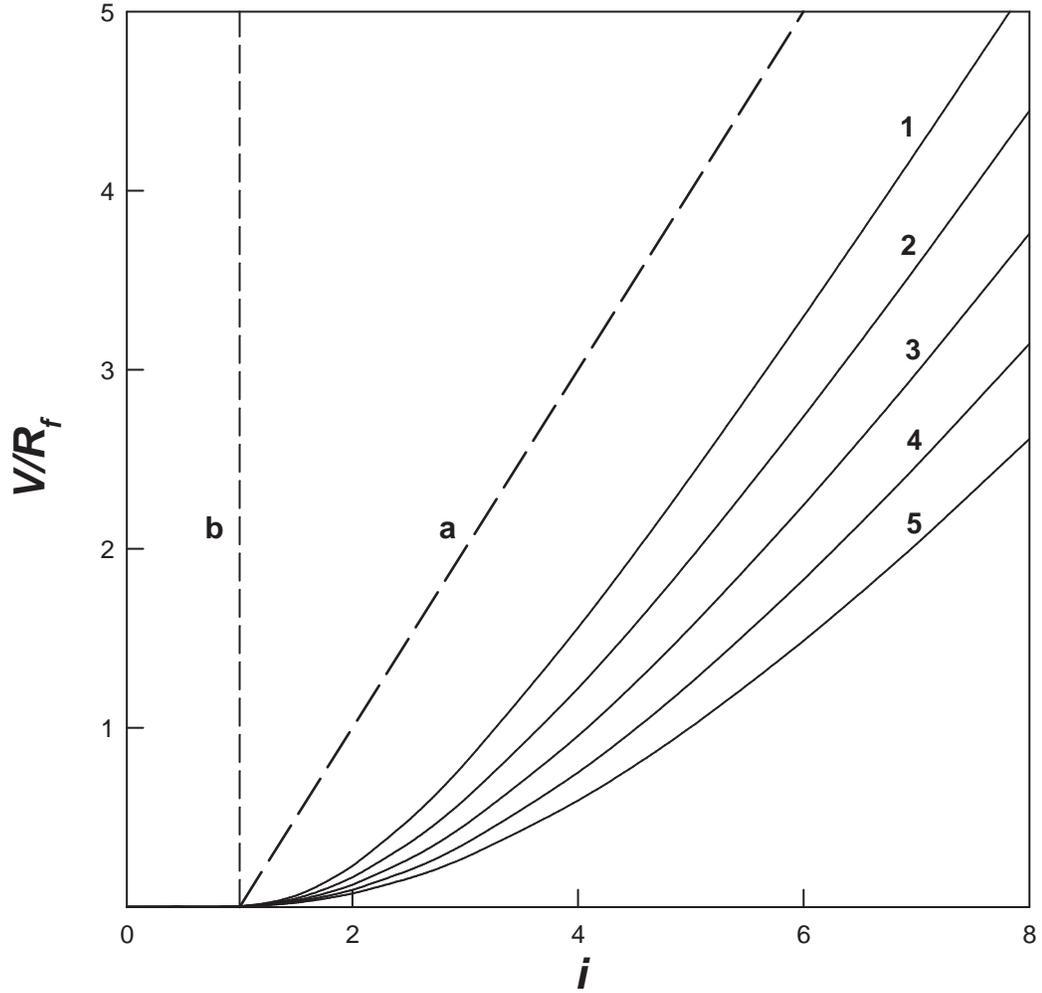}
\caption{Current-voltage characteristics of fractal superconducting
structures at $g=0$. Curve (1) corresponds to $D=1$, (2) - $D=1.25$, (3) - $%
D=1.5$, (4) - $D=1.75$, (5) - $D=2$. Dashed line $a$ corresponds to viscous
flow of a magnetic flux for the case of $\protect\delta $-like distribution
of the critical currents, and dashed line $b$ describes the superconductor
in a critical state.}
\label{figure7}
\end{figure}

Thus, accordingly with Equation (\ref{eq12}), for the $\delta $-like
distribution of the critical currents in the regime of viscous flow of the
magnetic flux the voltage across the superconductor is controlled by a
simple linear dependence $V=R_{f}\left( i-i_{c}\right) h\left(
i-i_{c}\right) $. The corresponding voltage-current characteristic is shown
in Fig.~\ref{figure7} by the dashed line $a$. In the same figure the
voltage-current characteristics of the superconductor in a still simpler
approximation of the critical state model is represented, too (the dashed
curve $b$). In this case the response on any external action, which results
in the appearance of an electric field in a hard superconductor, is the
flow of the current that is equal to the critical one $i=i_{c}$,
regardless of the value of the voltage across the sample. (The dimensionless
critical current is equal to unity due to the normalization chosen above, $%
i\equiv I/I_{c}$)

For the fractal distribution of the critical currents the situation changes
radically, because now the vortices are being broken away in a wide range
of transport currents. We shall consider the case of the exponential 
distribution of the cluster areas ($g=0$), where the current-carrying
capability of the superconductor is maximum. After substitution of the 
distribution of Eq.~(\ref{eq10}) into Eq.~(\ref{eq12}) followed by 
integration, the voltage across the sample can be expressed through the
function of cumulative probability of Eq.~(\ref{eq9}) 
\[
V=R_{f}\int_{0}^{i}F\left( i^{\prime }\right) di^{\prime } 
\]
Upon integration we get 
\begin{equation}
V=R_{f}\left[ i\exp \left( -\left( \frac{2+D}{2}\right) ^{\frac{2}{D}+1}i^{-%
\frac{2}{D}}\right) -\left( \frac{2+D}{2}\right) ^{\frac{2+D}{2}}\Gamma
\left( 1-\frac{D}{2},\left( \frac{2+D}{2}\right) ^{\frac{2}{D}+1}i^{-\frac{2%
}{D}}\right) \right]  \label{eq13}
\end{equation}
In the limiting cases of $D=1$ and $D=2$, Equation (\ref{eq13}) can be 
simplified.

For Euclidean clusters ($D=1$) the voltage across the superconductor has
the form 
\begin{equation}
V=R_{f}\left[ i\exp \left( -\frac{3.375}{i^{2}}\right) -\sqrt{3.375\pi }%
\text{erfc}\left( \frac{\sqrt{3.375}}{i}\right) \right]  \label{eq14}
\end{equation}
where the complementary incomplete gamma function is expressed through the
complementary error function $\Gamma \left( 1/2,z\right) =\sqrt{\pi }$erfc$%
\left( \sqrt{z}\right) $.

For clusters of maximum fractality ($D=2$), after substitution of the
representation for the complementary incomplete gamma function $\Gamma
\left( 0,z\right) =-%
\mathop{\rm Ei}%
\left( -z\right) $, Equation (\ref{eq13}) for the voltage takes its final
form 
\begin{equation}
V=R_{f}\left[ i\exp \left( -\frac{4}{i}\right) +4%
\mathop{\rm Ei}%
\left( -\frac{4}{i}\right) \right]  \label{eq15}
\end{equation}
where $%
\mathop{\rm Ei}%
\left( -z\right) $ is the exponential integral function.

\begin{figure}[tbp]
\epsfbox{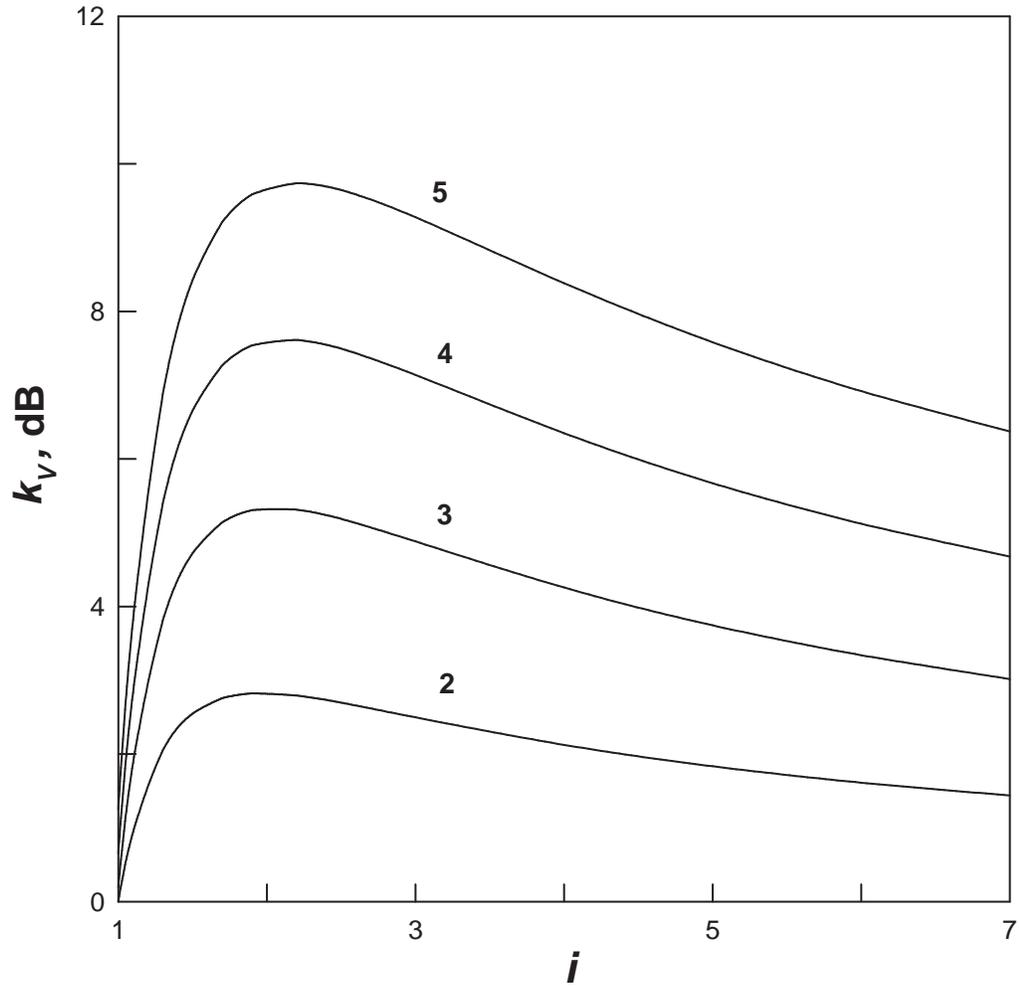}
\caption{Attenuation factor of dissipation at different values of the
fractal dimension. Curve (2) corresponds to $D=1.25$, (3) - $D=1.5$, (4) - $%
D=1.75$, (5) - $D=2$.}
\label{figure8}
\end{figure}

In Fig.~\ref{figure7} there are shown the voltage-current characteristics of
a superconductor containing fractal clusters of a normal phase. For all
values of the fractal dimension there is a noticeable voltage drop starting
from the transport current of $i=1$ that coincides with the value found
earlier from the critical current distribution of Eqs.~(\ref{eq5}) and (\ref
{eq9}) for the resistive transition current. Equations (\ref{eq14}) and (\ref
{eq15}) describe the dependencies of the voltage on the transport current
for the limiting values of the fractal dimension. Whatever the geometric
probability properties of the normal phase clusters may be, the
voltage-current characteristics of a superconductor with these clusters will
lie in the region confined by these dependencies (curves 1 and 5 in Fig.~\ref
{figure7}). As is seen from the figure, the fractality significantly reduces
the electric field induced by magnetic flux motion. In Fig.~\ref{figure8}
the dependencies of the attenuation factor of dissipation 
\[
k_{V}\equiv 20\log \frac{V\left( D=1\right) }{V\left( D\right) }\text{, \ \
dB} 
\]
on the transport current at various values of the fractal dimension are
presented. The decrease in the electric field with increasing fractal
dimension is especially appreciable in the range of currents $1<i<3$, where
the pinning gain also has a maximum (Fig.~\ref{figure6}). Both of these
effects have the same nature, since their reason lies in the peculiarities
of the fractal distribution of the depinning currents. As is seen from Figs.~%
\ref{figure3} and \ref{figure4}, an increase in the fractal dimension leads
to considerable spreading of the tail of the distribution $f=f\left(
i\right) $. This means that more and more clusters of small sizes, which can
best trap the magnetic flux, are being involved in the game. As a result,
the density of vortices broken away from pinning centers by the Lorentz
force decreases, so the smaller part of magnetic flux can flow creating a
lower electric field. In turn, the smaller the electric field, the smaller
energy is dissipated when the transport current passes through the sample.
So the decrease in heat evolution, which could cause a transition into a
normal state, leads to an increase in the current-carrying capability of the
superconductor containing such fractal clusters.

Thus, Figs.~\ref{figure5}-\ref{figure8} clearly demonstrate the most
important result: the fractality of the normal phase clusters prevents
destruction of the superconductivity by the transport current, and thereby
enhances the current-carrying capability of the superconductor. The fractal
clusters essentially affect the dynamics of the magnetic flux trapped in the
superconductor. The crucial change of the depinning current distribution
caused by increasing the fractal dimension of the clustes is the
reason of this effect. The fractality of the boundaries of the normal phase
clusters intensifies pinning and slows down destruction of the
superconductivity by the transport current. This point provides principally
new possibilities for increasing the critical currents of composite
superconductors by optimizing their geometric morphological properties.

This work is partially supported by Russian Foundation of Fundamental
Researchs (Grant No~02-02-17667).

\newpage

\newpage

\newpage

\newpage

\newpage

\newpage

\newpage

\newpage

\end{document}